\documentclass[fleqn,usenatbib]{mnras}

\usepackage{newtxtext,newtxmath}
\usepackage[T1]{fontenc}
\DeclareRobustCommand{\VAN}[3]{#2}
\let\VANthebibliography\thebibliography
\def\thebibliography{\DeclareRobustCommand{\VAN}[3]{##3}\VANthebibliography}

\usepackage{graphicx}	
\usepackage{amsmath}	
\usepackage{txfonts}
\usepackage{xcolor}
\usepackage{hyperref}
\usepackage{orcidlink}
\usepackage{placeins}
\usepackage{natbib,twoopt}

\title[The EECS determination]{An empirical determination of the Cosmic Shoreline}

\author[P. Meni-Gallardo et al.]{
P. Meni-Gallardo,$^{1,2}$ 
and E. Pall\'e$^{1,2}$
\\
$^{1}$Instituto de Astrof\'{i}sica de Canarias (IAC), 38205 La Laguna, Tenerife, Spain\\
$^{2}$Departamento de Astrof\'{i}sica, Universidad de La Laguna (ULL), 38206 La Laguna, Tenerife, Spain\\
}

\date{Accepted 2026 June 15. Received 2026 June 5; in original form 2026 March 3}

\pubyear{2026}

\begin{document}
\label{firstpage}
\pagerange{\pageref{firstpage}--\pageref{lastpage}}
\maketitle

\begin{abstract}

The cosmic shoreline concept was introduced to separate planets with atmospheres from those without, by relating the cumulative X‑ray and extreme‑ultraviolet (XUV) instellation (integrated over the planet’s lifetime) to the planetary escape velocity, using the Solar System planets to anchor the empirical relation. The exoplanet community has since attempted to refine the cosmic shoreline to provide a consistent ranking or prioritisation tool for exoplanet observations – i.e., to quickly identify which small planets are most likely to have retained an atmosphere and therefore merit expensive follow‑up with facilities such as JWST or the upcoming ELTs. Here, we use an empirical approach to refine the Cosmic Shoreline concept. In particular, we used the data provided by the ExoAtmospheres database, and the NASA Exoplanet Archive, along with solar system data. We reconcile limitations in the classical shoreline definition by anchoring our Empirical Exoplanet Cosmic Shoreline (EECS) simultaneously to Mars, GJ 9827 d, L 98–59 d, $\rm GJ~3090~b$, and Pi Mensae c (all having tentative atmospheric detections).  The EECS exhibits a significantly steeper slope than previously theorized, while consistently categorising Solar System moons and dwarf planets according to their atmospheric properties. Applied to planets orbiting M dwarfs, the EECS suggests that a larger fraction retain atmospheres than predicted by classical models, but incorporating revised XUV fluence histories for low-mass M dwarfs reveals severe atmospheric vulnerability.
We finally identify high-priority targets for the JWST Rocky Worlds survey and future ELTs observations based on their EECS positioning and Transmission/Emission Spectroscopy Metrics. 
\end{abstract}



\section{Introduction}

Exoplanetary science has seen a vast development over the past three decades, leading to a large diversity of confirmed planetary systems. With each new progress, the field is focussing more and more on the search and characterisation of small worlds: sub-Neptunes, super-Earths, and Earth-like planets. 
This pursuit is driving ambitious space missions and the construction of extremely large ground-based telescopes (ELTs), with temperate, Earth-sized habitable zone planets as key targets.
Following JWST's successful launch, the next milestone is the ELTs' first light. By the time the first ELTs operate (2030s), current and upcoming missions such as TESS \citep{Ricker2015}, CHEOPS \citep{Benz2021}, and PLATO \citep{Rauer2014}, plus ground-based surveys, should have identified and characterised the most promising potentially habitable planets. If JWST is not able to do it for a few selected targets, high-resolution spectrographs on ELTs, already in advanced design, will be the first instruments capable of probing habitable zone planet atmospheres \citep{Palle2025andes}. Their role is to go beyond mass and radius, exploring atmospheric structure, composition, surface conditions, and potential biomarkers \citep{DesMarais2002}.

M dwarfs dominate the solar neighbourhood \citep{henry18, reyle21}, and their smaller size facilitates the detection of lower-mass planets. However, theory suggests that planets around M stars are highly vulnerable to atmospheric loss \citep{Luger2015, Wordsworth2015, Dong2018}, raising concerns about their long-term habitability under intense stellar radiation \citep{Schaefer2016, Owen2016}. Currently, we lack even the basic knowledge of whether M dwarf rocky habitable zone planets possess atmospheres. The initial JWST results are at best ambiguous \citep{Lustig2023, Moran2023}, which led to a major public DDT programme to address this issue \citep{Redfield2024}. This highlights the need for optimal strategies and target selection to maximise the science return of these expensive facilities. 

Atmospheric escape is governed primarily by photoevaporation, the heating of the upper atmosphere by stellar XUV radiation, which drives a hydrodynamic outflow that can remove large fractions of the gaseous envelope. This process is particularly effective for close‑in planets, which receive high stellar insolation. Planets can acquire atmospheres through different channels. Primary atmospheres are captured directly from the protoplanetary nebula and are typically H/He‑rich; these are easily lost for low‑mass planets. Secondary atmospheres are outgassed from the planetary interior (e.g., via volcanism) or delivered by impacts, and are typically composed of heavier volatiles (e.g., CO$_2$, H$_2$O, N$_2$). The distinction is important because secondary atmospheres are more resilient to escape due to their higher mean molecular weight. Thus, a planet’s atmospheric state reflects its formation history – accretion location, migration, and the availability of volatiles in the protoplanetary disk – and the subsequent “nurturing” or stripping by stellar activity over billions of years.

To tackle this problem, \citet{zahnle2017cosmic} introduced the concept of the “cosmic shoreline” to separate planets that have retained an atmosphere from those that have not. The underlying idea is that atmospheric escape is driven by the cumulative X‑ray and extreme‑ultraviolet (XUV) radiation received from the host star over the planet’s lifetime, while gravity forces work to retain the atmosphere. The relevant parameters are therefore the cumulative XUV instellation $I_{XUV}$=($a_\oplus^2$/$a_\star$)$^2$($L_\star$/$L_\odot$)$^{0.4}$ (the total stellar XUV flux integrated over time) and the planetary escape velocity $v_{esc} $=$ \sqrt{(2GM_p) \text{/} (R_p)}$. \citet{zahnle2017cosmic} proposed a relation of the form $I_{XUV}$ = $K v_{esc}^4$ + $C$, which becomes linear in log‑log space: $log_{10} I_{XUV}$ = $4 log_{10}v_{esc}$ + $D$. The slope of 4 arises from the energy‑limited escape framework, where the mass‑loss rate scales as the XUV flux divided by the escape velocity, and the total mass lost over time scales as the integrated XUV flux divided by the escape velocity. 

Given the lack of information on exoplanet atmospheres at the time, they tried to constrain the parameters that define this concept by using the most reliable data they had: the Solar System planets. The free parameter $D$ (i.e. the zero point) was originally fixed by assuming that Mars marks the boundary between worlds with and without atmospheres in the Solar System. They proposed that a line could be drawn that separates whether the planets are able to retain an atmosphere or not. Using this method they classified the Solar System objects in terms of their atmosphere and extended it to the whole template of exoplanets. 

Other recent works have built on the original cosmic shoreline idea. \cite{ji2025cosmicshorelinerevisitedmetric} revisited the concept of cosmic shoreline, focussing on atmospheric retention of rocky planets, by incorporating hydrodynamic escape models for various atmospheric compositions. They computed time-integrated atmospheric loss across a range of stellar and planetary parameters, considering realistic stellar XUV evolution. Their analysis demonstrates that the shoreline is probably not a sharp boundary, but a broad transition zone shaped by non-linear escape physics and an initial volatile inventory. \cite{pass2025receding} also revised estimates of atmospheric retention for rocky exoplanets orbiting mid-to-late M dwarfs by incorporating updated stellar activity lifetimes and XUV emission histories. Using recent observations of stellar rotation and magnetic activity, the authors find that these stars remain in the saturated regime far longer than previously assumed, leading to significantly higher cumulative XUV fluence. As a result, many of the known planets shift above the cosmic shoreline, implying that only the most massive rocky planets are likely to retain an atmosphere. 
Other authors \citet{ih2025, Berta-Thompson2025}, have tried statistical approaches for detailed small plane characterisation and to provide refined frameworks for prioritising observational programmes focused on exploring small planet atmospheres.

Despite the complexity proposed by these authors, which we can not still test with the available data, in this paper, we provide an alternative, empirical definition of the cosmic shoreline, based on the cumulative knowledge to date of exoplanet atmospheric observations and solar system exploration data. 

\section{Data}

In this paper, we make use of data from the IAC community ExoAtmospheres database\footnote{\url{https://research.iac.es/proyecto/exoatmospheres/index.php}}. The ExoAtmospheres database is a repository of all exoplanet atmospheric detections, collecting information on the properties of the planet, chemical species detected, atmospheric properties, and related literature. Up to now 58 different chemical species have been detected among 274 explored planets, ranging from ultra-hot Jupiters to temperate rocky worlds. 

We downloaded the ExoAtmospheres database and created an up-to-date list of which planets have empirical detections of an atmosphere, that takes into account planet type; atmospheric species detections, featureless spectra retrievals, and albedos. While it is not possible to infer unequivocally from the ExoAtmospheres data that a planet has no atmosphere (as a non-detection could always be related to poor data or an atmospheric signal below the detection capability of our instrumentation), it is possible to make a list of those planets that do have an atmosphere via the statistically significant detection of a given chemical species.

The stellar and planet parameters used by ExoAtmospheres are imported from the Exoplanet Encyclopaedia\footnote{\url{https://exoplanet.eu/home/ database}}, and if some parameters are missing, we import them from the NASA Exoplanet Archive\footnote{\url{https://exoplanetarchive.ipac.caltech.edu}}. 
These data include stellar parameters (such as magnitude and temperature), exoplanet parameters (such as masses and orbital parameters) and discovery/characterisation data (such as published radial velocity curves, photometric light curves, images, and spectra). 

\begin{figure*}[h!]
    \centering
    \includegraphics[width=15cm]{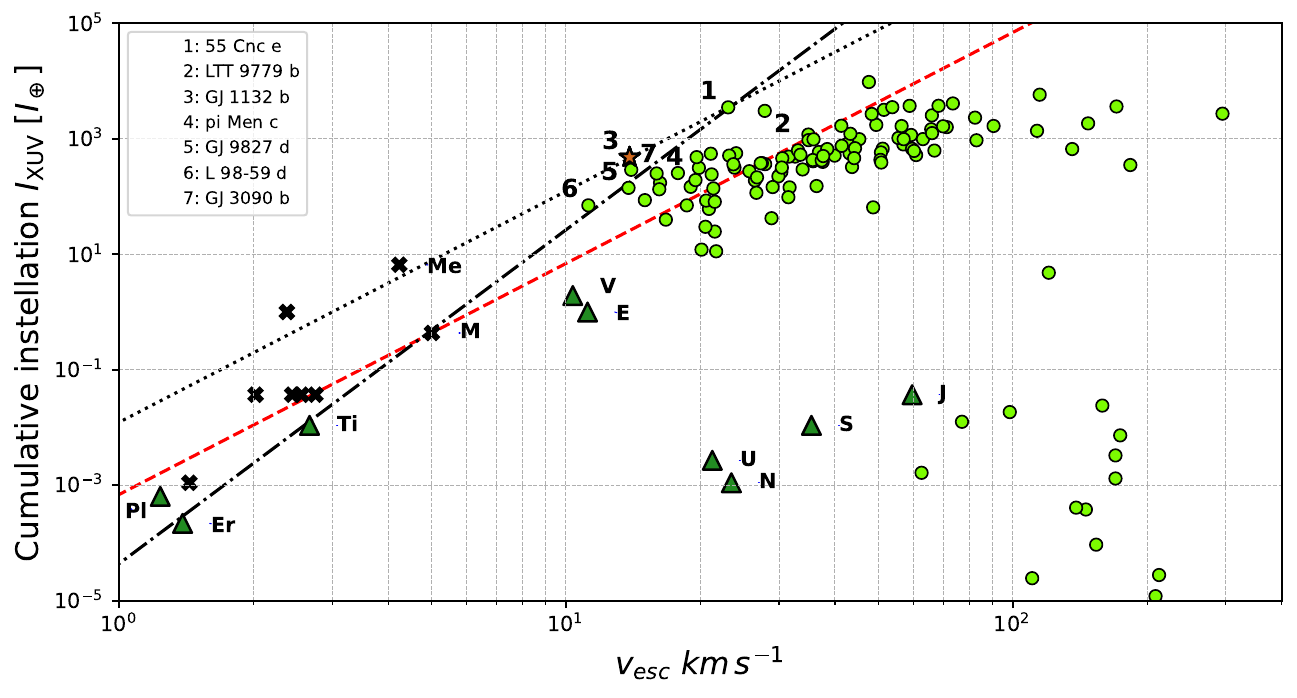}
    \caption{The Cosmic Shoreline. Light‑green dots are all known exoplanets with reported atmospheric detections from the ExoAtmospheres database, noting that some of these detections are tentative or remain contested in the literature (e.g., GJ 1132 b). Solar System objects are marked as black crosses or green triangles if they host an atmosphere. The red dashed line is the cosmic shoreline as defined by \citet{zahnle2017cosmic}.  The black dotted is the same classical line, preserving the slope, but taking as zero point the exoplanet $\mathrm{55~Cnc~e}$. The black dashed-dotted line is our newly proposed Empirical Exoplanet Cosmic Shoreline (EECS), which makes use of both Mars and several exoplanet atmospheric detections to define the slope and zero-point value. The numbers highlighted in the inset show the names of the seven planets discussed in the definition of the EECS.}
    \label{fig:New_csl}
\end{figure*}

\section{Methods}
\label{sec:meth}
\subsection{Classical Cosmic Shoreline definition}

The "classical" definition of the cosmic shoreline by \cite{zahnle2017cosmic} provides a non-linear relation between the cumulative XUV instellation and the escape velocity, which becomes linear in the logarithmic space. The expected slope for this shoreline is 4, but the zero point remains unconstrained. 
\cite{zahnle2017cosmic} considered energy‑limited escape. In this framework, the mass‑loss rate from a planet's atmosphere is $\dot{M} \propto F_{\mathrm{XUV}} R_p$ /($ G M_p$), where $F_{\mathrm{XUV}}$ is the incident XUV flux. Integrating over time yields a total mass loss proportional to the cumulative instellation $I_{\mathrm{XUV}}$ times $R_p$/($G M_p$). Since $v_{\mathrm{esc}}$ = $\sqrt{2GM_p\text{/}R_p}$, one obtains $M_{\mathrm{lost}} \propto I_{\mathrm{XUV}} $/$ v_{\mathrm{esc}}^2$. Setting the condition that a planet retains an atmosphere if the initial volatile inventory exceeds the total possible loss leads to the critical relation $I_{\mathrm{XUV}} \propto v_{\mathrm{esc}}^4$. The constant of proportionality remains unconstrained by the energy‑limited theory alone, so \cite{zahnle2017cosmic} used Mars to set the zero point: they assumed that Mars is the transition object in our Solar System, having a very thin atmosphere but still located just on the retention side of the shoreline.

However, when confronted with the latest atmospheric data from exoplanets, this proposed cosmic shoreline does not fit the observations (see Figure \ref{fig:New_csl} and the discussion in the following sections). In this work, we explore the possibility of using the known population of exoplanets to better constrain this zero-point and slope. But first we need to establish which planets have statistically significant atmospheric detections.

Throughout this work the $I_{XUV}$ instellation was computed using the equations from \cite{zahnle2017cosmic}, except for the analysis of the M dwarf planet population, in Figures \ref{fig:newfig3} and \ref{fig:habitableplanets}, where for stellar hosts with $\rm M_s < 0.35 M_{\odot}$ we used the $I_{XUV}$ individual values from \cite{pass2025receding}.

\subsection{Which planets have an atmosphere?}

There are several planets whose combination of high cumulative XUV instellation, low escape velocity and atmospheric explorations, place them in the bordeline of the classical cosmic shoreline, but seven of them stand out over the rest. The first of them is $\mathrm{GJ~1132~b}$ \citep{Berta_Thompson_2015}, a warm Earth-sized planet with a radius of $1.16\,\mathrm{R_\oplus}$, a mass of $1.65\,\mathrm{M_\oplus}$, an equilibrium temperature of $589\, \mathrm{K}$ and a period of $1.63\, \mathrm{days}$ orbiting a M3.5 dwarf. Although ExoAtmospheres lists a detection of HCN in its atmosphere \citep{Swain2021}, this detection has not held up and more recent results \citep{2023_May_1132, Palle2025crires, Bennett2025} report non-detections of HCN, featureless spectra or establish stringent constraints for the existence of its atmosphere. Due to this uncertainty, in this work, we do not consider GJ 1132b to have an unequivocally detected atmosphere or to define the limit of the EECS.

Another planet in the borderline is $\mathrm{LTT~9779~b}$ \citep{Pearson_2019}, an ultra-hot Neptune, with a radius of $0.42\,\mathrm{R}_J$, a mass of $0.09\,\mathrm{M}_J$, an equilibrium temperature of $2000\,\mathrm{K}$ and a period of $0.79 \, \mathrm{days}$ orbiting a G8 star. $\mathrm{LTT~9779~b}$ has different atmospheric species detections such as FeH, CO$_2$ and H$_2$O \citep{2023_Edwards} and upper limits for OH \citep{2025_zhou}. Although $\mathrm{LTT~9779~b}$ is a solid placement for the cosmic shoreline limit at high irradiation regimes, it is not the best option because it causes several other planets with detected atmospheres to fall into the non-atmospheric retention region.

$\mathrm{55~Cnc~e}$ \citep{mcarthur2004detection} is a hot super-Earth with a radius of $1.95\,\mathrm{R_\oplus}$, a mass of $8.58\,\mathrm{M_\oplus}$, an equilibrium temperature of $1996\,\mathrm{K}$ and an orbital period of $0.74\, \mathrm{days}$ orbiting a G8V star. It has a very high cumulative XUV instellation, low escape velocity, and the recent discovery of a secondary atmosphere \citep{Renyu2024}. Thus, $\mathrm{55~Cnc~e}$ is currently a solid candidate to establish the empirical limit between exoplanets retaining and non-retaining atmospheres. We note, however, that the source of the variability and composition of the atmosphere of 55 Cnc e is still unclear \citep{Zilinskas2025}, likely the product of a unique extreme environment as an ultrashort-period planet. Extrapolating from it to the entire population of exoplanets, the majority of which are not in such extreme configurations, is thus a big jump. Thus, we decided to also exclude 55 Cnc e as a possible anchor to define the EECS. Still, whether we choose 55 Cnc e or not, the EECS relationship described in this work will not change, as we will discuss in the next section.

With all previous considerations, we use $\rm GJ~9827~d$, $\rm L~98-59~d$, $\rm GJ~3090~b$ and $\rm Pi~Mensae~c$ as our EECS anchors. These four planets have in common the tentative detections of possessing an atmosphere -- $\rm GJ~9827~d$ \citep{Roy2023, Piaulet2024}, $\rm L~98-59~d$ \citep{Banerjee2024, Gressier2024}, $\rm GJ~3090~b$ \citep{2025Ahrer}, $\rm Pi~Mensae~c$ \citep{Garcia2021}. Our aim is to test whether a monotonic relation (covering as wide a range of XUV irradiation as possible) is sufficient, with our present level of knowledge, to define a boundary, which indeed seems to be the case.
We describe this process in the following section.

\subsection{The Empirical Exoplanet Cosmic Shoreline definition}

In Figure \ref{fig:New_csl}, we represent all known exoplanets with detected atmospheres, based on the ExoAtmospheres database, and we also include the planets and the major moons of the Solar System. The classical cosmic shoreline \citep{zahnle2017cosmic} is marked with a red broken line, which clearly leaves out a large number of planets with strong atmospheric detections ($\mathrm{LTT~9779~b}$ for example). In the figure, we have two alternative ways of defining a new Empirical Exoplanet Cosmic Shoreline. 

The first is to preserve the slope of the classical Cosmic Shoreline, using $\mathrm{55~Cnc~e}$ to define the zero point. This is shown in Figure \ref{fig:New_csl} as a dotted line. The line leaves Mercury just above it on the non-retention region. However, this line is problematic as many Solar System moons that have no atmospheres would fall below it, on the atmospheric retention region. Choosing as zero point any of the other six planets discussed in section 3.2 causes a similar problem. 
 
The second way is by using both Mars and the four selected planets in section 3.2 to define a boundary line. This is based on the original conception of Mars as the edge between planets with or without atmosphere in our Solar System. This modifies the classical cosmic shoreline, leading to a new slope of $\approx$5.77. Taking the form of:
\begin{equation}
    \log_{10}(I_{XUV})\approx 5.77 \log_{10}(v
    _{esc})-4.35
\end{equation}

This second boundary line (dash-dotted in the Figure \ref{fig:New_csl}), which we will name from now on the Empirical Exoplanet Cosmic Shoreline (EECS), seems to do extremely well in separating known atmospheric detections from bodies without significant atmospheres. Notably, the extremely irradiated super-Earth 55 Cnc e also falls exactly on top of this line, providing additional support for this boundary to remain valid at high cumulative XUV instellation values

Moreover, using $\rm Pi~Mensae~c$ (and {\it{de facto} $\mathrm{55~Cnc~e}$}) has the advantage that both host stars are G-type, and thus the cumulative insolation flux received by the planets should not be underestimated \citep{pass2025receding}. In this sense, \citet{france2025semiempiricalestimatesolareuv} constrained the EUV activity of G-type stars using a semi-empirical approach, claiming that the oldest stars can show large uncertainties in the integrated XUV radiation, since it is dominated by the early high-activity phase. In particular, they point out that our knowledge of the early stages of stellar evolution is still limited and that their sample is not large enough to establish a universal evolutionary law. Thus, most of the uncertainty is introduced during those early stages, when the XUV flux is highest and its level and duration are less well constrained. At later ages, although the instantaneous XUV flux is relatively low, the uncertainty associated with the early phases does not disappear but is carried into the integral. Future refinements of individual target's XUV irradiance histories will likely influence the exact location of the EECS. However, given the estimated age of 55 Cnc e ($\approx$10.2 Gyr), we expect this correction to be small. The steeper slope of the EECS is also in line with the findings of \citet{Berta-Thompson2025}, which approach the definition of a cosmic shoreline from a statistical point of view. The steeper slope of the EECS (5.77) is in line with the statistical inference of \citet{Berta-Thompson2025}, who derived a slope of $5.9^{+0.61}_{-0.43}$ in the escape‑velocity dimension using a probabilistic 3D framework.

At even lower cumulative XUV instellation, Titan and the dwarf planet Eris in our own Solar System also fall on top of this EECS line (Figure \ref{fig:New_csl}). Titan definitely possesses an atmosphere \citep{kuiper1944titan, kunde1981c4h2, horst2017titan}, while the case of Eris is curious. Eris has a thin, transient atmosphere, due to the periodic sublimation of its surface ices of methane and nitrogen along its orbit \citep{Hofgartner2019}. 

This EECS boundary line is also much more consistent with the properties of Solar System objects, as it places Europa, Callisto, Ganymede, Haumea and Triton, which have no significant atmospheres (certainly not detectable at interstellar distances), in the non-retention region. Only Pluto, which has a mass and radius comparable to Eris, and an established atmosphere \citep{Elliot1989}, is in disagreement with the EECS, but there is no way to reconcile Pluto, Mars as a boundary planet, and the know exoplanet atmosphere population with a simple shoreline. Perhaps the case of Pluto/Eris illustrates that a simple relationship might not hold for the very tenuous atmospheres displayed by small (exo-)solar system bodies.

\subsection{The Atmosphere Retention Metric}

Once our new EECS has been established, it is worth confronting it with more generalised metrics reported in the literature. \cite{pass2025receding} proposed the concept of and Atmosphere Retention Metrics (ARM), also based on the cumulative XUV instellation that planets suffer and the escape velocity, and described as:
\begin{equation}
 \label{eq:airgarm}
    ARM\equiv 4\log_{10}(v_{esc})-\log_{10}(I_{XUV})-ZP_{Mars}
\end{equation}

Where $v_{esc}$ is the escape velocity, $I_{XUV}$ is the cumulative XUV instellation and $ZP$ is an arbitrary zero point, which they chose to be located at the Mars ARM value. 

\cite{pass2025receding} proposed this metric for planets orbiting stars with $\rm M_* \leq 0.35 M_\odot$, and reconstructed their individual cumulative stellar XUV instellation histories, taking into account all the radiation sources that these planets have undertaken during their lifetime. Here, we make use of these new metrics in our figures, but changing the slope value to 5.77 and taking as our zero point the ARM value of the EECS, such as:
\begin{equation}
 \label{eq:ours}
    ARM\equiv 5.77\log_{10}(v_{esc})-\log_{10}(I_{XUV})-4.35
\end{equation}
 
\begin{figure*}
    \centering
    \includegraphics[width=21cm]{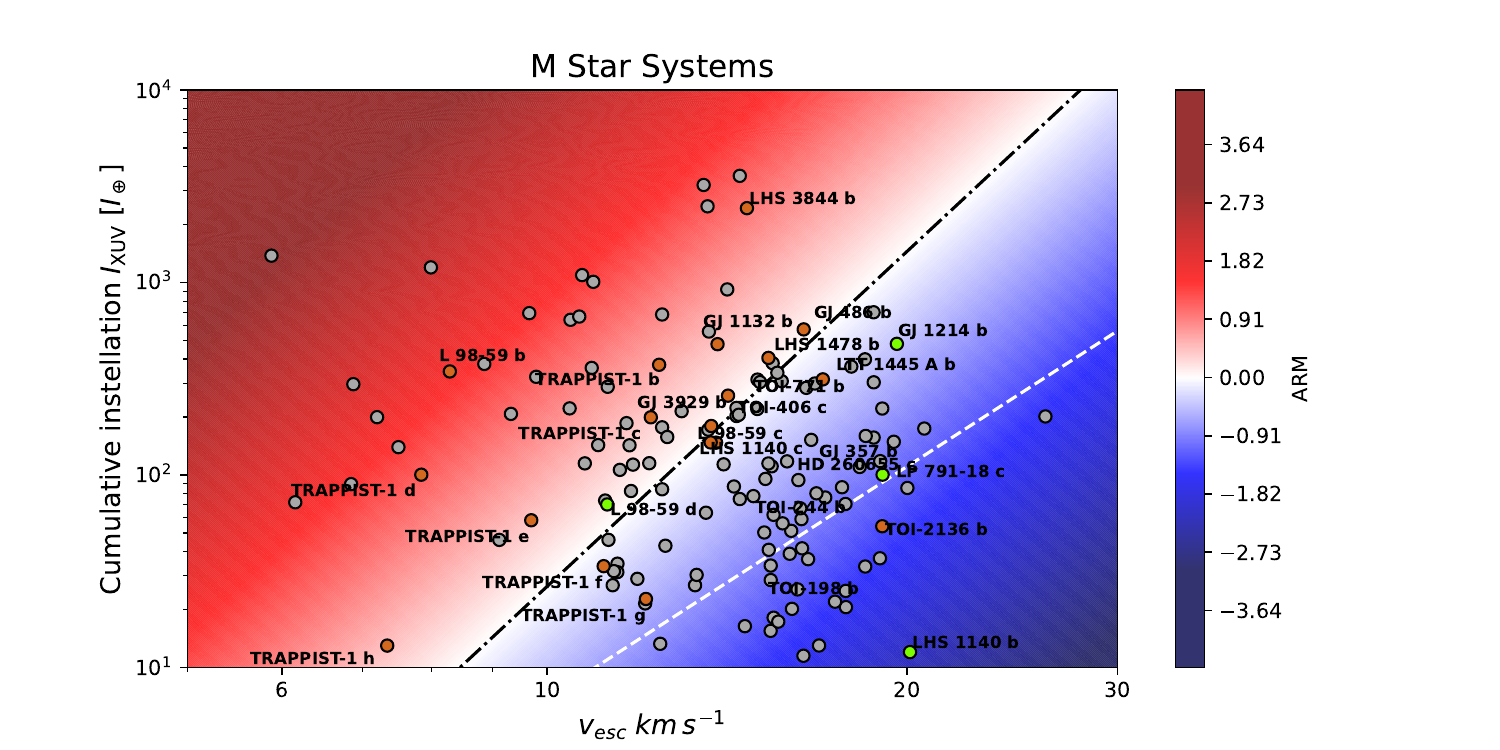}
    \caption{Cosmic Shoreline for all known planetary systems orbiting M-dwarf-type stars. The black dashed-dotted line is our newly proposed Empirical Exoplanet Cosmic Shoreline (EECS) and the white-dashed line is the classical cosmic shoreline, same as in Figure \ref{fig:New_csl}. The gray points mark the systems whose atmospheres have not been studied; the orange ones mark the planets with reported strong non-detections; and the light-green ones mark the planets with a (tentative) atmosphere detection. 
    \label{fig:Comp_ARM}}
\end{figure*}

\section{Results and Discussion}

\subsection{Implications for M dwarf planets population}

\begin{figure*}
    \centering
    \includegraphics[width=21cm]{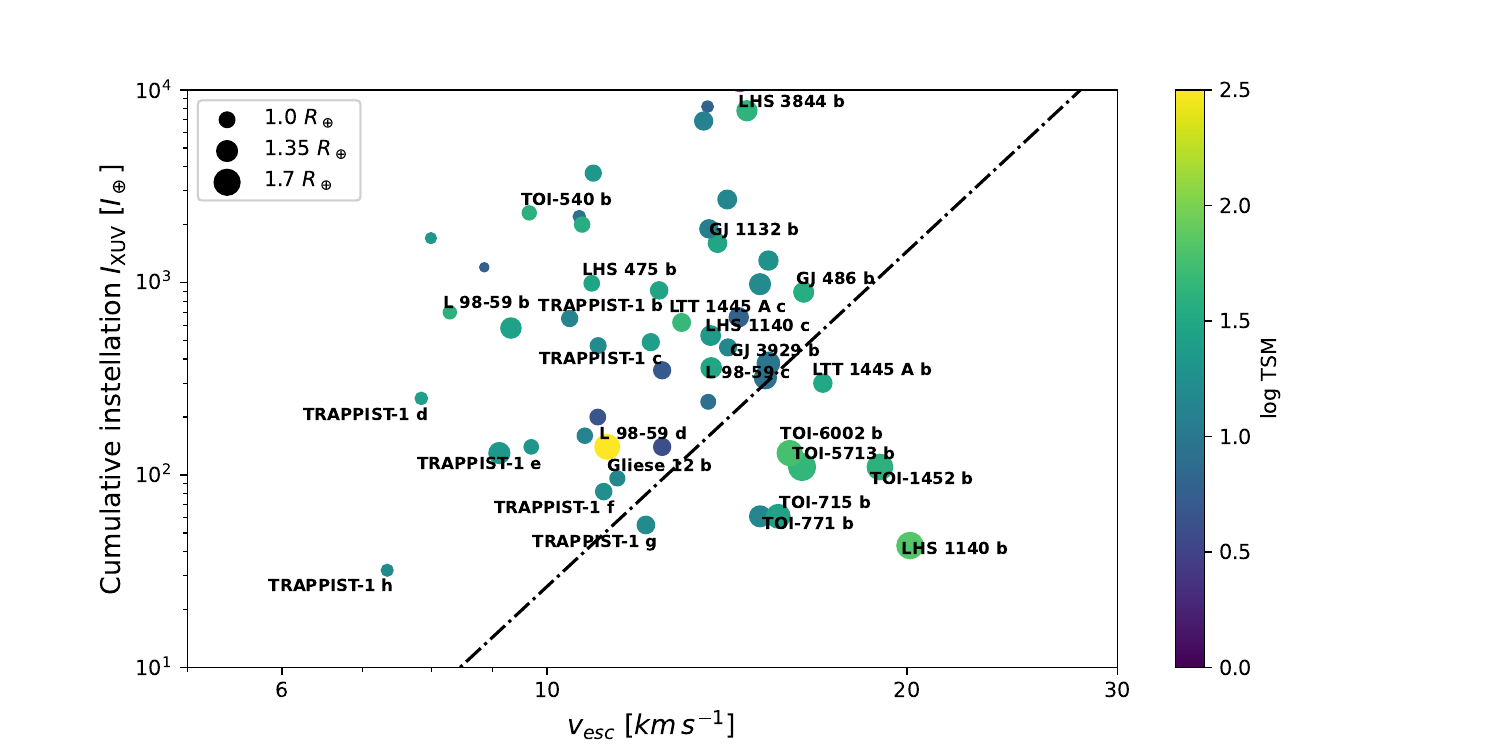}
    \includegraphics[width=21cm]{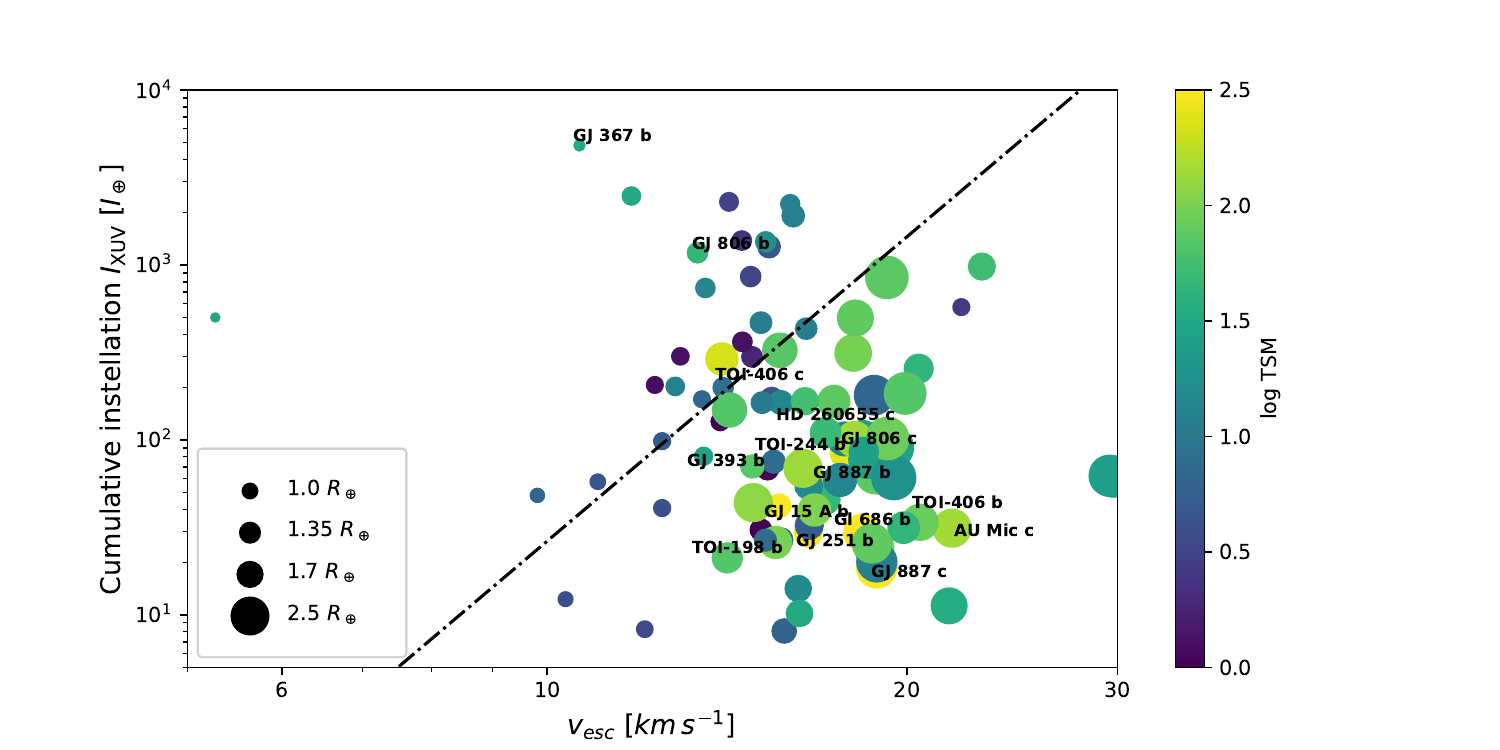}  
    \caption{Empirical Exoplanet Cosmic Shorelinefor all known planetary systems orbiting M-dwarf-type stars. The top panel shows the planets around stellar hosts with $\rm M_s < 0.35 M_{\odot}$ and the $I_{XUV}$ individual values of \citet{pass2025receding}. The lower panel shows planets with stellar hosts with $\rm M_s > 0.35 M_{\odot}$ and the  $I_{XUV}$ values calculated following \citet{zahnle2017cosmic}. Overplotted are the names of the planetary systems closer than 15 pc to Earth. In the top panel we also name TOI-1452 b, TOI-6002 b, TOI-5713 b and TOI-715 b as the fall below the EECS; and GJ 3929 b and TOI-771 b as they are the selected JWST DDT targets. On the bottom panel we also mark JWST DDT targets TOI-198 b, TOI-406 c, TOI-406 b, HD 260655 c, and TOI-244 b.  The colour scale codes the TSM metric and the symbol size the planetary radius.}
    \label{fig:newfig3}
\end{figure*}

\begin{figure*}
    \centering
    \includegraphics[width=21cm]{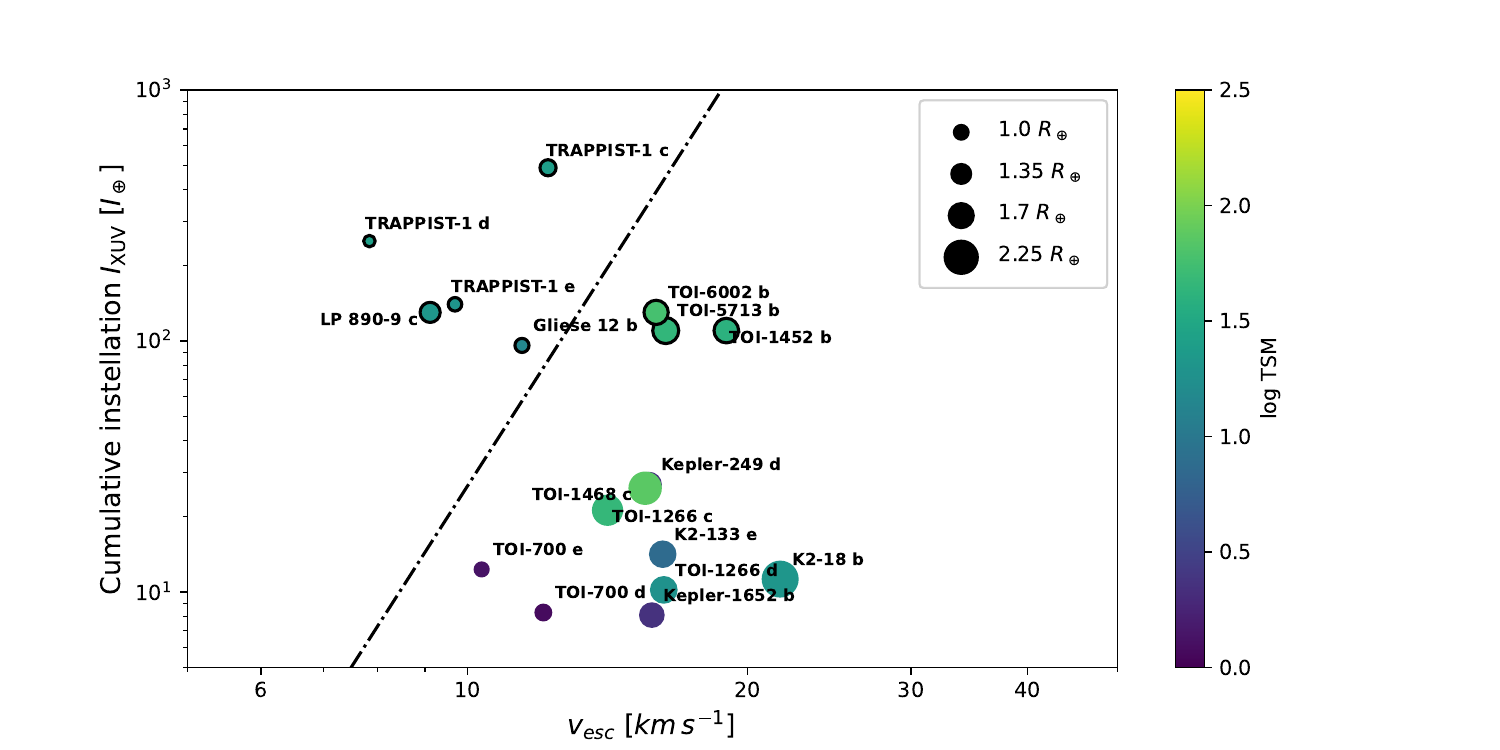}
    \caption{Empirical Exoplanet Cosmic Shoreline for habitable planets orbiting M-dwarf-type stars. A $v_{esc}$ vs $I_{XUV}$ diagram of all known planets with $\rm R < 2.5 R_{\oplus}$ and equilibrium temperature between 250 and 350 K. Planets with  $I_{XUV}$ values from \citet{pass2025receding} are plotted with a solid line around the symbol.  The color scale codes the TSM metric and the symbol size the planetary radius.}
    \label{fig:habitableplanets}
\end{figure*}

As discussed in the introduction, practically all habitable zone small rocky planets accessible for atmospheric characterisation with the JWST or the upcoming ELTs orbit around M dwarfs \citep{Palle2025andes}. Thus, we focus on planets around these planetary systems in order to assess the chances of this population to retained atmospheres.  

Figure \ref{fig:Comp_ARM} shows the location of all planets orbiting M-dwarf stars, in a $v_{esc}$ vs $I_{XUV}$ diagram, calculated following \citet{zahnle2017cosmic}, over plotted to the ARM values following Equation  \ref{eq:ours}. Planets that have already been observed in search for signs of an atmosphere, with the most precise instruments to date, are marked with an orange point, these are: TRAPPIST-1$~$b \citep{2023Lim, 2023Greene}, TRAPPIST-1$~$c \citep{2025Rathcke, 2025Radica}, 
TRAPPIST-1$~$e \citep{2025Berardo}, TRAPPIST-1$~$f \citep{Krishnamurthy2021, 2025Berardo}, TRAPPIST-1$~$g \citep{2019Wakeford, 2025Berardo}, TRAPPIST-1$~$h \citep{2022Garcia}; L$~$98-59$~$b \citep{2025Bello, 2024Scarsdale}, L$~$98-59$~$c \citep{2025Barclay, 2023Zhou}, GJ$~$1132$~$b \citep{May2023, Swain2021, Palle2025crires}; GJ$~$486$~$b \citep{2023Ridden-Harper}; LTT$~$1445$~$A$~$b \citep{2025Bennet, 2023Diamond}; LHS$~$3844$~$b \citep{2020Diamond},LHS$~$1478$~$b \citep{august2025hot} and TOI-2136$~$b \citep{2022Kawauchi}. Only four planets around M dwarfs have (very) tentative evidences of possessing an atmosphere, and are marked in light-green: LHS$~$1140$~$b \citep{2021Edwards, Cadieux2024}, GJ$~$1214$~$b \citep{Orell2022, 2024Everett, 2025Kazumasa}, LP$~$971-18$~$c \citep{2025RoyLP} and L$~$98-59$~$d \citep{2024Banerjee, 2024Gressier}. 

Figure \ref{fig:Comp_ARM} proposes a much more positive scenario for M dwarf planets atmospheric retention than the classical shoreline definition. More than half of the known planets population move from outside to inside the atmospheric retention region. In particular, three of the four tentative detections are above the classical shoreline definition, but all fall below our EECS (or very close to the boundary in the case of L$~$98-59$~$d), in the atmospheric retention area. However, the results shown in Figure \ref{fig:Comp_ARM}, may be too optimistic as M dwarfs cumulative $I_{XUV}$ is not easy to calculate. 

\cite{pass2025receding} argued that the insolation received by planets orbiting around M dwarfs has been systematically underestimated, implying that significantly fewer planets orbiting such stars would be able to retain an atmosphere, especially those orbiting around stellar hosts with $\rm M_s < 0.35 M_{\odot}$. The top panel of Figure \ref{fig:newfig3} shows only the planetary systems orbiting such small-mass M dwarf hosts, but this time using the cumulative insolation values reported in \cite{pass2025receding}.  Their $I_{XUV}$ calculations take into account the contributions from the pre-main-sequence phase, the saturation timescale, and the flare bursts that these stars are expected to have experienced.  We note that ideally, we should also obtain the modified $I_{XUV}$ for Mars and $\mathrm{55~Cnc~e}$ in order to recalculate the zero point of the ARM. However, given that both planets orbit G-type stars, the differences should be negligible within the errors, and we have ignored this correction in this work. The bottom panel of Figure \ref{fig:newfig3} shows the planet orbiting hosts with $\rm M_s > 0.35 M_{\odot}$ and using the $I_{XUV}$ estimates of \citet{zahnle2017cosmic}.

Figure \ref{fig:newfig3} also offers information on the size of the planets and on their potential for atmospheric characterisation by means of TSM metrics (Transmission Spectroscopy Metrics; \cite{Kempton18}). We find that the majority of planets orbiting low-mass M dwarfs, those easier to characterise and generally with the smallest radii, fall above the EECS. Only seven systems would be able to retain their atmospheres based on our current estimate of the EECS. In contrast, there are a large number of planets below the EECS for the most massive M dwarf hosts. However, most have radii equal to or larger than about 1.7 $\rm R_\oplus$. 

Of the seven targets below the EECS in the top panel of Figure \ref{fig:newfig3}, only $\rm TRAPPIST-1~g$, $\rm LHS~1140~b$ and $\rm LTT~1445~A~b$ have been searched for atmospheres, in all cases reporting non-detections (featureless spectra), except for $\rm LHS~1140~b$, for which there is a water detection \citep{2021Edwards}. In the case of $\rm TRAPPIST-1~g$, observations were made by HST WFC3 and STIS \citep{2019Wakeford, 2025Berardo}, while for $\rm LTT~1445~A~b$ there are ground-based observations at low-resolution with Magellan II/LDSS3C \citep{2020Diamond} and from space with HST WFC3 \citep{2025Bennet}. Follow up observations of these objects with JWST and the ELTs would be highly valuable to test the exact limits and validity of the EECS proposed here.

\subsection{Implications for the JWST {\it Rocky Worlds} survey}

Currently, we lack the basic knowledge of whether small planets around M dwarfs, especially those orbiting in the habitable zone of their stars, have atmospheres at all. The first results from the JWST observations are ambiguous at best \citep{Lustig2023, Moran2023}, and to try to solve this issue a very large public DDT programme, named {\it Rocky Worlds}\footnote{https://rockyworlds.stsci.edu/index.html}, has recently been approved \citep{Redfield2024}. 

Given that the {\it Rocky Worlds} initiative is adopting the strategy of measuring secondary eclipse depths as a means to detect an atmosphere, in Figures \ref{fig:Appendix1} and \ref{fig:Appendix2} we reproduce Figures \ref{fig:newfig3} and \ref{fig:habitableplanets}, respectively, but with the Emission spectroscopy metrics in the color axis.

The {\it Rocky Worlds} programme has already selected its first targets. According to our EECS, the first two targets, $\rm LTT~1445~A~c$ and $\rm GJ~3929~b$, with already on-going observations for the latter, would not have retained an atmosphere (Figure \ref{fig:newfig3} top). In the case of $\rm GJ~3929~b$, results have been published reporting a non-detection, in line with the EECS predictions \citep{xue2025jwstrockyworldsddt}. The recent additions to the programme, $\rm LHS~1140~b$ and $\rm LTT~1445~A~b$ have larger radii, but are much better choices, with $\rm LHS~1140~b$ well within the atmospheric retention regions and $\rm LTT~1445~A~b$ placed close to the transition region but also on the retention side. To illustrate a broader set of targets, we highlight a few additional examples included in the programme: TOI-198 b, TOI-406 c, TOI-771 b, HD 260655 c, and TOI-244 b to highlight a few examples. According to the EECS presented in this work, these new target are reasonable selections and expected to have retained an atmosphere.

For future extensions of the Rocky Worlds programme beyond its current targets, $\rm TOI-1452~b$, $\rm TOI-6002~b$, $\rm TOI-5713~b$, $\rm TOI-771~b$ and $\rm TOI-715~b$ would be the best (and so far the only) options for observing small ($R <2R_\oplus$) planets with reasonable TSM values that also lie within the atmospheric retention region of low‑mass M dwarfs (top panel of Figure \ref{fig:newfig3}). For planets orbiting more massive M dwarfs, $\rm GJ~15~A~b$, $\rm GJ~887~b$, $\rm TOI-406~b$, $\rm GJ~251~b$, $\rm TOI-406~b$ $\rm TOI-198~b$, and $\rm GJ~393~b$ are the most interesting small candidates with favourable TSM. In the particular case of $\rm TOI-5713~b$, the cumulative XUV instellation given in \cite{pass2025receding} is only an upper limit due to the high stellar activity level. This would move TOI-5713 b further down in Figure \ref{fig:Comp_ARM}, making it an even more favourable target for atmosphere retention. However, active stars present their own challenges regarding the feasibility of follow up with JWST \citep{Lim:2023}. Unfortunately, exploring the atmospheres of the majority of these targets would require a prohibitive amount of JWST observing time, and thus their characterisation might be left for the ELTs or future space missions.

One of the major limitations in establishing a definitive EECS is the lack of clear boundaries outside the Solar System, i.e. the lack of established non-existence of exo-atmospheres, as high mean molecular weight atmospheres for many planets do not fall within the detectability limits of JWST. However, repeated observations of specific targets could help determine very stringent upper limits to the placement of the EECS. 

One of the major limitations in establishing a definitive EECS is the lack of clearly established “non‑existence” of exoplanet atmospheres. By “non‑existence” we mean that repeated, high‑sensitivity observations (e.g., JWST thermal phase curves or transmission spectra) have placed such stringent upper limits on the atmospheric pressure that even a tenuous, high‑mean‑molecular‑weight atmosphere can be ruled out. For example, the innermost planets of the TRAPPIST‑1 system have now been observed with JWST thermal phase curves, showing a very high day‑night temperature contrast that is incompatible with any significant atmosphere; TRAPPIST‑1 b is found to be essentially airless \citep{Gillon2026}. Similarly, for TRAPPIST‑1 d, \cite{Piaulet-Ghorayeb2025} reported strict limits that exclude secondary atmospheres above a few tens of millibars for many compositions. These examples demonstrate that it is becoming possible to empirically identify airless worlds near the shoreline. However, for most planets in the M‑dwarf population, the current data are still not deep enough to conclusively establish a non‑detection; this is why we argue that targeted campaigns to establish “non‑existence” near the EECS would be as valuable as positive detections.

In this sense, it would be important to also observe candidates near but above the EECS which have favourable TSM values, such as $\rm GJ~12~b$, $\rm L~98-59~b$ or $\rm GJ~486~b$. These solid "delimiting" non-detection planets might be as valuable as any eventual atmospheric detection in order to more solidly define the EECS, and in turn help us identify the most robust criterion for target selection to substantially enhance the scientific return of future observations. As an example, \citet{Bennett2025} very recently published a joint analysis of four transits and one emission dataset, all from JWST, and placed very stringent constraints on the possibility of an atmosphere of GJ 1132 b. Empirically establishing GJ 1132 b, or any other planet near the EECS, undoubtedly as bare rocks would be a major step into defining the true limits of the EECS (see Figure \ref{fig:New_csl}), and might be well worth a campaign of repeated observations with the JWST.

\subsection{Implication for habitability searches}

Beyond determining the existence or lack of atmospheres for M dwarf planets, another important question is the exploration of small planets in the habitable zone, including the search for possible biomarkers \citep{DesMarais2002, harmanschwieterman2015}. In Figure \ref{fig:habitableplanets} we plot all the confirmed transiting planets orbiting M dwarfs, with $\rm R <2.5R_\oplus$, and with an equilibrium temperature between 250 and 350 K. Currently, only 14 planets meet these criteria. One of these planets is K2-18~b, a sub-Neptune planet, proposed as a possible hycean planet \citep{Madhusudhan2023}, which has a confirmed atmosphere but also an on-going intensive debate on its overall nature and detected atmospheric species \citep{Schmidt2025, Madhusudhan2025, Hu2025}. Three more sub-Neptune planets TOI-1468 c, TOI-1266 b and TOI-1266 c have similar sizes and TSM to K2-18 b and can be relatively easily explored with JWST in the near future. 

For truly small earth-size planets, the c, d, and e planets in the Trappist1 system and Gj-12 b  are the best, most accessible targets, but unfortunately all lie on the atmosphere-less part of the EECS. Only TOI-700 e and d are expected to possess an atmosphere, although their TSM metrics make them prohibitive targets for JWST. Thus, based on our EECS results, sub-Neptune planets emerge as the only possible targets for biomarker searches with currently available instrumentation. 

One planet deserves special attention. Given its size, its location in the habitable zone, its TSM metrics, and its proximity to our definition of the EECS, Gj-12 b \citep{Kuzuhara2024, Turner2025, brady2025earthlikedensitytemperateearthsized} appears a fundamental planet for JWST to explore in depth, to determine both the possibility of the planet hosting an atmosphere and the exact limits of the EECS.

\subsection{Limitations and potential evolution of the EECS}

The EECS, like the classical shoreline, is presented here as a single, sharp dividing line. In reality, the transition between planets with and without atmospheres is likely to be fuzzy, due to differences in initial volatile inventory, stochastic impacts, and variations in stellar XUV history. Recent work by \cite{Ji2025} explicitly shows that varying the initial volatile content from 0.01\% to 1\% of the planetary mass can significantly shift the critical instellation threshold, blurring any unique shoreline. \cite{Berta-Thompson2025} introduced a statistical framework for a “3D cosmic shoreline” that accounts for this fuzziness by inferring a probability of atmosphere retention.

If a clear counter‑example to the EECS were to arise (such as a planet that clearly falls in the “non‑retention” region of Figure 1 but nonetheless possesses a robustly detected atmosphere), our framework would need to evolve. One possibility would be to adopt a probabilistic version of the shoreline, where the boundary is not a line but a logistic regression surface, as proposed by \cite{Berta-Thompson2025}. Another direction would be to include a third parameter that captures the planet’s likely volatile inventory, e.g., its bulk density or inferred envelope mass fraction \citep{Ji2025}. Finally, the emerging population of ultra‑short‑period rocky planets that appear to retain atmospheres (e.g., 55 Cnc e, and the very recent detection of a thick volatile atmosphere on TOI‑561 b by \cite{Teske2025}) suggests that high‑temperature secondary atmospheres, possibly sustained by magma‑ocean outgassing, may follow a different retention criterion than H/He‑dominated primary atmospheres. Indeed, TOI‑561 b orbits an old ($\approx 10$ Gyr) thick‑disk star and has a density of only $4.3 \pm 0.4$ $g$/$cm^{3}$, implying a significant volatile layer \citep{Teske2025}, although the atmosphere has not been directly detected. In the EECS diagram, TOI‑561 b lies slightly above the line (see Figure 1), meaning that our current EECS would predict a bare rock. If further observations confirm a robust atmosphere, this would indicate that the EECS slope or zero point needs adjustment, or that a separate “ultrahot secondary atmosphere” shoreline is required. We therefore encourage continued high‑precision observations of borderline targets, as they will be the ones to refine (or refute) the EECS.

\section{Conclusions}

In this study, we have established a new empirical definition of the cosmic shoreline, the EECS, by leveraging atmospheric detections across exoplanetary systems and Solar System objects. Our approach improves upon the classical shoreline by incorporating empirical data from both the Solar System (Mars) and exoplanets with tentative atmospheric detections (GJ 9827 d, L 98–59 d, $\rm GJ~3090~b$, and Pi Mensae c). The EECS aligns with additional extreme cases such as 55 Cnc e, which exhibits a secondary atmosphere despite extreme irradiation. Thus, the resulting EECS provides a significantly improved fit to the observational data. The slope of 5.77 reflects a steeper dependence on escape velocity than previously theorised.

Our analysis reveals critical implications for planets orbiting M dwarfs, prime targets for atmospheric characterisation with JWST and future ELTs. When applying the EECS with classical \(I_{\mathrm{XUV}}\) estimates, over half of known M-dwarf planets shift into the atmospheric retention zone—a more optimistic outlook than the classical shoreline suggests. However, incorporating updated cumulative XUV fluence histories for low-mass M dwarfs ($\rm M_* < 0.35\,M_\odot$) from \cite{pass2025receding} dramatically reduces the fraction of planets capable of retaining atmospheres. For these systems, only seven small planets ($\rm R \lesssim 1.7\,R_\oplus$) fall securely below the EECS, emphasising the vulnerability of rocky worlds to prolonged stellar activity. This refinement underscores the need to prioritise targets such as TOI-1452 b and TOI-715 b for the {\it Rocky Worlds} survey, as they combine favourable TSM with EECS-based retention potential. Conversely, in-depth observing campaigns for planets near but above the EECS could provide essential constraints through non-detections.  

Looking ahead, the EECS offers a robust framework for guiding the search for habitable environments. For temperate, Earth-sized planets in the habitable zone, our results suggest that TOI-700 e and d are promising atmospheric candidates, although their low TSM values challenge near-term characterisation. Conversely, TRAPPIST-1 planets c to e and GJ 12 b likely lie in the atmosphere-depleted region, reducing their astrobiological priority. GJ 12 b, however, lies closer to the transition region and its properties make it a prime target for JWST in-depth observations. Future efforts must focus on expanding empirical validations of the EECS, particularly through high-precision observations of borderline candidates and systems with well-constrained XUV histories. Such data will not only refine the shoreline’s slope and zero-point, but also optimise the scientific return of next-generation telescopes in the quest to identify habitable worlds.  

\section*{ DATA AVAILABILITY}
The data underlying this article are public and are available in NASA Exoplanet Archive ( \url{https://exoplanetarchive.ipac.caltech.edu/}) and in ExoAtmospheres from IAC (\url{https://research.iac.es/proyecto/exoatmospheres/index.php}).

\section*{\it Acknowledgements}
\it The authors wish to thank Mrs Weronika \L oboda for useful discussions during this research project. We acknowledge financial support from the Agencia Estatal de Investigaci\'on of the Ministerio de Ciencia e Innovaci\'on MCIN/AEI/10.13039/501100011033 and the ERDF “A way of making Europe” through project PID2021-125627OB-C32, and from the Centre of Excellence “Severo Ochoa” award to the Instituto de Astrofisica de Canarias.



\bibliographystyle{mnras}
\bibliography{citations} 




\begin{appendix}
\section{Additional Figures}
\label{Sec:Appendix}

\begin{figure*}
    \centering
    \includegraphics[width=18cm]{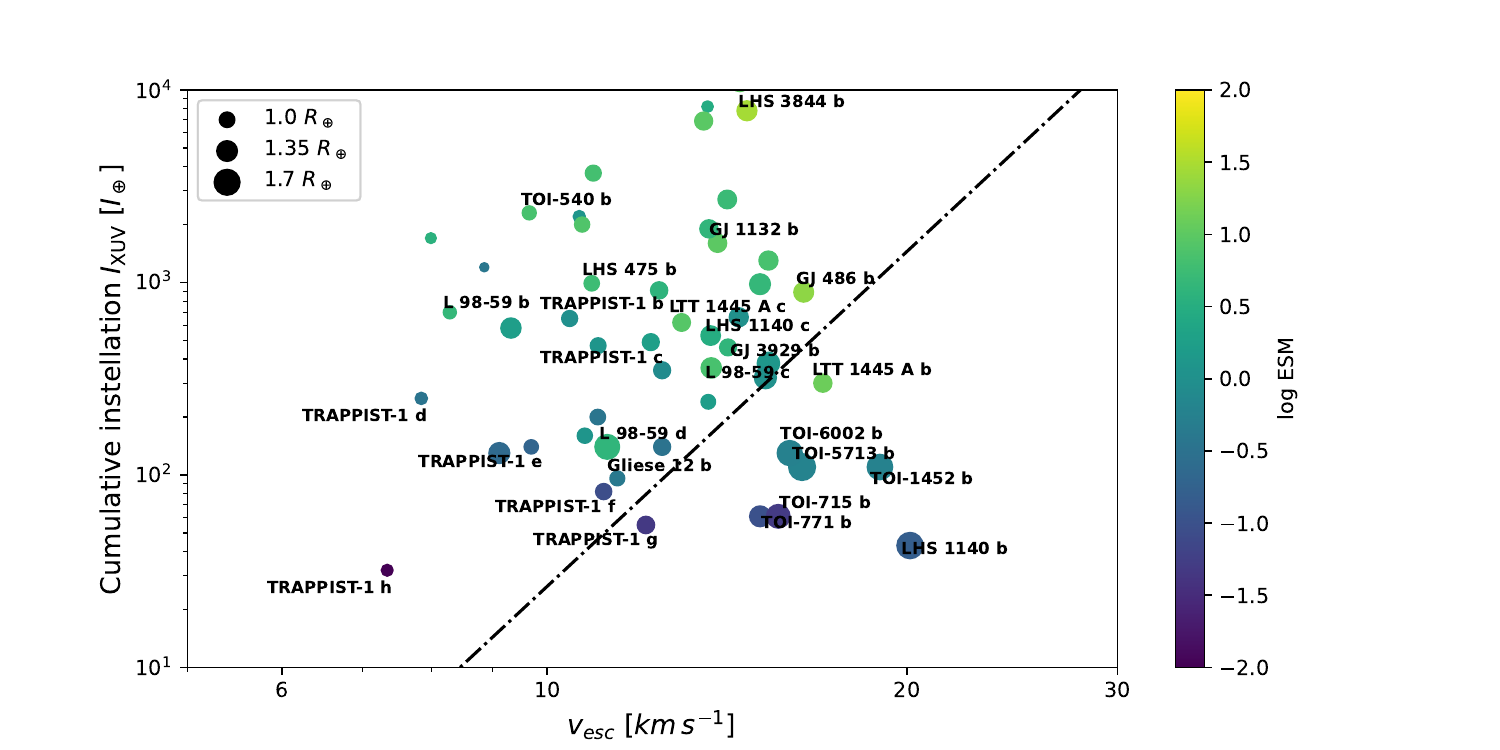}
    \includegraphics[width=18cm]
    {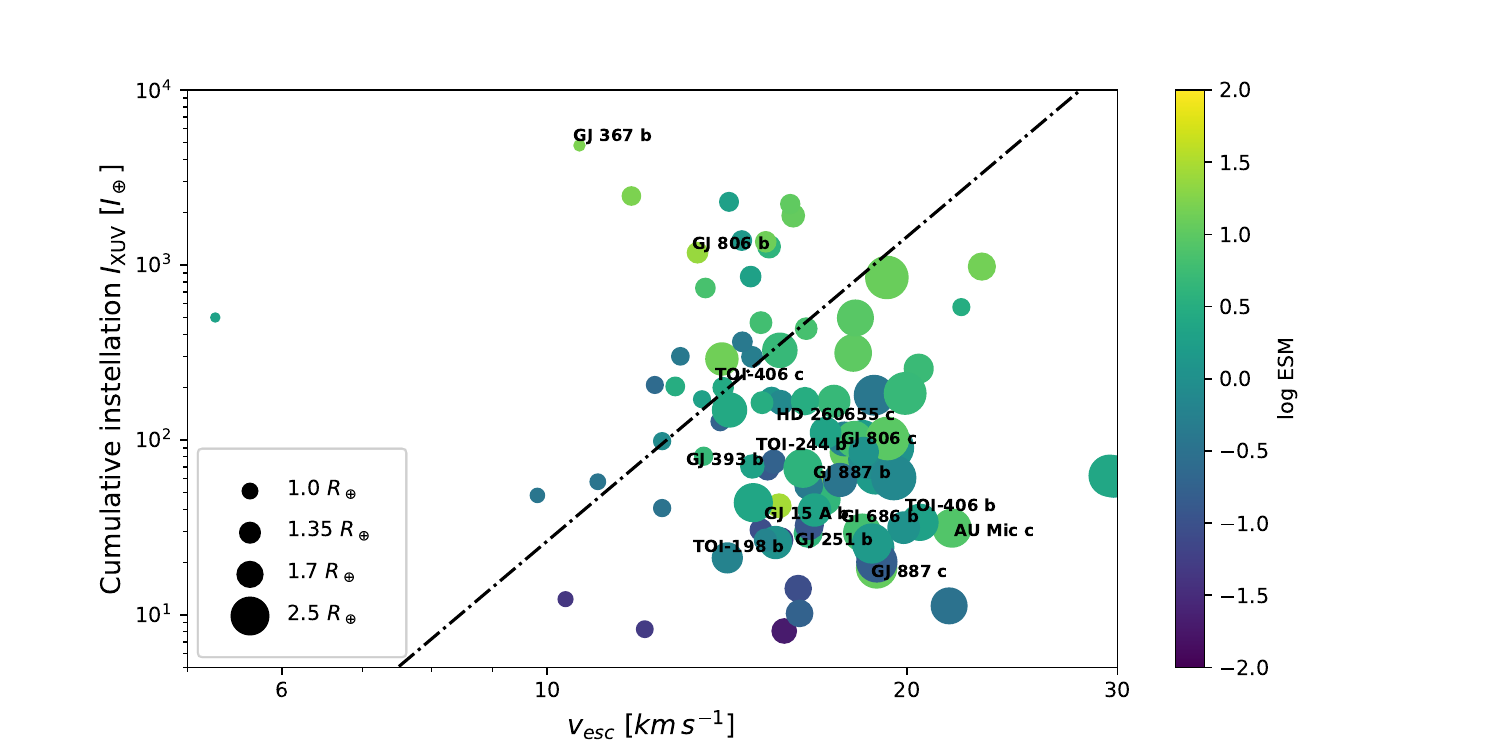}
    \caption{Same figure as Figure \ref{fig:newfig3} but with the ESM in the colorbar instead of the TSM.}
        \label{fig:Appendix1}
\end{figure*}

\begin{figure*}
    \centering
    \includegraphics[width=18cm]{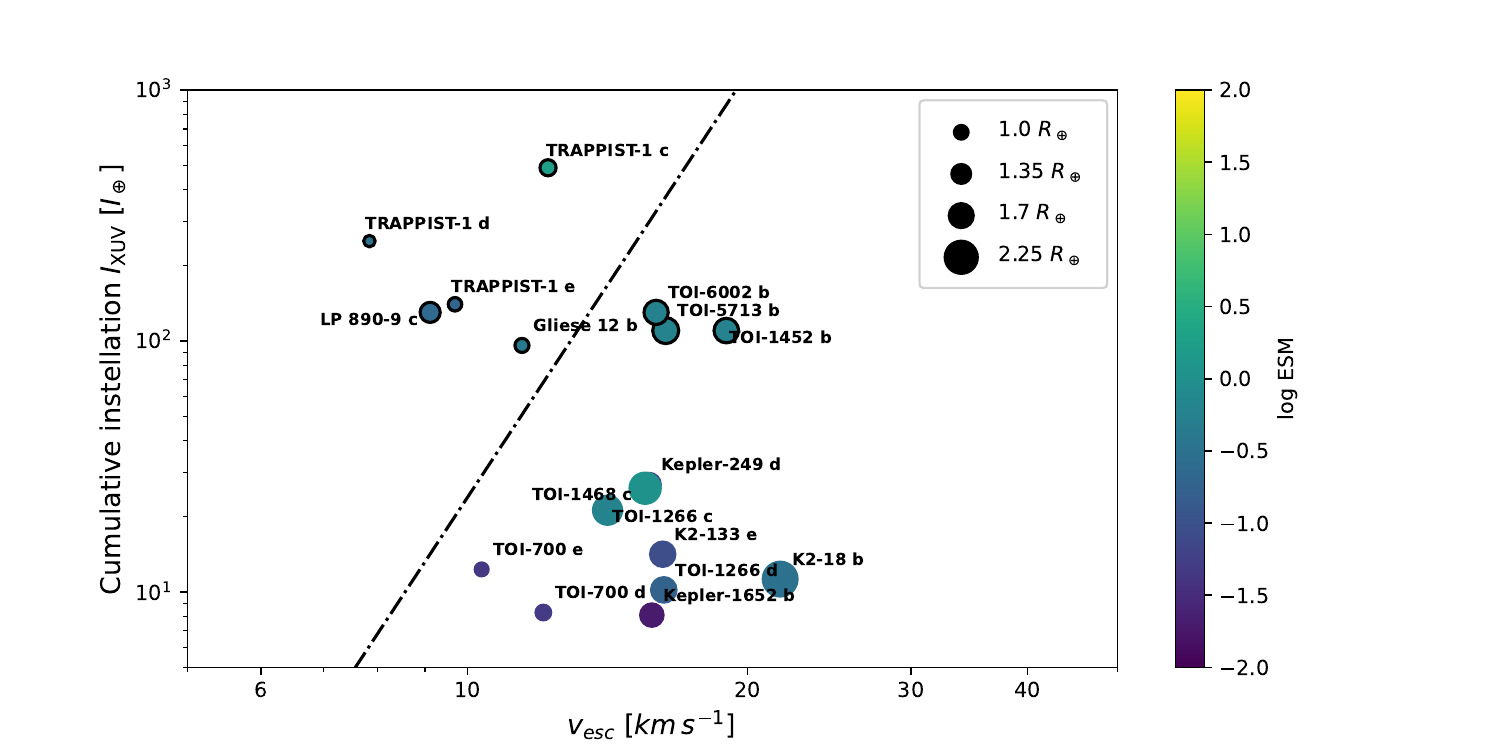}
    \caption{Same figure as Figure \ref{fig:habitableplanets} but with the ESM in the colorbar instead of the TSM}
    \label{fig:Appendix2}
\end{figure*}

\end{appendix}


\bsp	
\label{lastpage}
\end{document}